\documentclass[11pt,letterpaper]{article} 

\usepackage{iftex}
\ifPDFTeX
  \usepackage[T1]{fontenc}
  \usepackage[utf8]{inputenc}
  \usepackage{lmodern}
\else
  \usepackage{fontspec}     
\fi

\usepackage[margin=1in]{geometry}
\usepackage[final]{microtype}
\usepackage{setspace}   
\usepackage{amsmath,amssymb,amsthm,mathtools}
\numberwithin{equation}{section}

\usepackage{graphicx}
\graphicspath{{figs/}} 
\usepackage[labelfont=bf,font=small]{caption}
\usepackage{booktabs}
\usepackage{siunitx}
\usepackage{makecell} 
\usepackage{tikz}
\usepackage{tabularx}

\usepackage{enumitem}
\setlist{nosep}

\usepackage[hidelinks]{hyperref} 
\usepackage[nameinlink,capitalize,noabbrev]{cleveref}

\usepackage[numbers,sort&compress]{natbib}   
\usepackage{authblk}
\newcommand{\corrfoot}{\thanks{Corresponding authors: \texttt{bakker@mit.edu}; \texttt{vaccaro@mit.edu}}}

\title{Evaluating Human-AI Safety: A Framework for Measuring Harmful Capability Uplift}
\author[1]{Michelle Vaccaro\corrfoot}
\author[1]{Jaeyoon Song}
\author[1]{Abdullah Almaatouq}
\author[1]{Michiel A. Bakker$^*$}
\affil[1]{Massachusetts Institute of Technology, Cambridge, USA}
\date{January 28, 2026}

\newcommand{\keywords}[1]{\par\smallskip\noindent\textbf{Keywords:} #1}

\begin{document}
\maketitle

\begin{abstract}
\noindent
Current frontier AI safety evaluations emphasize static benchmarks, third-party annotations, and red-teaming. In this position paper, we argue that AI safety research should focus on human-centered evaluations that measure harmful capability uplift—the marginal increase in a user's ability to cause harm with a frontier model beyond what conventional tools already enable. We frame harmful capability uplift as a core AI safety metric, ground it in prior social science research, and provide concrete methodological guidance for systematic measurement. We conclude with actionable steps for developers, researchers, funders, and regulators to make harmful capability uplift evaluation a standard practice.
\end{abstract}


\section{Introduction}
Recent advances in frontier AI models have dramatically expanded their capabilities. Systems that once struggled with basic question answering now produce runnable software code, integrate and interpret multiple forms of data (e.g., images, text, audio), and complete complex tasks at the level of domain experts. Consequently, their potential to benefit as well as harm society has expanded dramatically. In response, the AI safety community has developed an extensive toolkit of in-vitro evaluations---benchmarks for truthfulness, toxicity, bias, refusal consistency, jailbreak resistance, autonomy, and more \citep{Gehman2020-he, Lin2022-op, Rauh2022-cc, Chao2024-md, Cui2024-rm, Liu2023-rf}. These tests are fast, reproducible, and increasingly standardized, with each major model debut accompanied by a scorecard of headline metrics.

However, strong performance on static benchmarks often coexists with headline-grabbing failures in the wild \citep{El-Atillah2023-op, Nelken-Zitser2024-zs, Milmo2023-yo}. For example, frontier models that pass toxicity filters can still amplify extremist rhetoric when prompted creatively \citep{Gilbert2024-ig}. Empirically, improvements in scores on safety benchmarks track general capability scaling, leaving open the possibility of ``safetywashing''---relabeling raw performance improvements as safety progress \citep{Ren2024-se}. While some model evaluations involve people, most position humans as external judges rather than embedded actors. Researchers may recruit people to label outputs for harmfulness \citep{Bai2022-jp, Cheong2025-et, Grey2025-tk}, but they rarely assess how much harm people can cause when using the same model as a co-conspirator. Moreover, current static evaluation approaches cannot capture the harms that emerge through sustained human-AI interactions \citep{Ibrahim2024-fy}.

Given these limitations, we argue that AI safety research should incorporate human-centered evaluations that focus on measuring harmful capability uplift—the incremental change in a user's capacity to cause harm when assisted by frontier models. Evaluating harmful capability uplift shifts the focus of AI safety from ``Does the model ever emit dangerous content?'' toward ``Does the model meaningfully increase the harmful actions users can perform?'' 

Evaluating harmful capability uplift, however, is methodologically demanding. It requires experiments with human subjects that capture the dynamic, adaptive ways people incorporate suggestions from frontier models into their workflows. As highlighted by \citep{Ibrahim2024-fy}, human--computer interaction (HCI) research offers useful tools for AI safety research. For instance, researchers have for decades conducted user studies \citep{Lazar2017-fh}, run controlled experiments \citep{Carroll1997-ds}, and developed theory-driven models of augmentation \citep{Licklider1960-ip}; yet these methods have been used almost exclusively for benign applications—such as writing assistance or medical decision-making—rather than for assessing malicious scenarios. Malicious tasks introduce distinct methodological challenges, including adversarial objectives, hidden ground truth, and serious ethical constraints on ``live-fire'' trials.

To position harmful capability uplift as a standard practice in AI safety evaluation, we structure our analysis around four key contributions: First, we examine the three main pillars of today's safety evaluation---static benchmarks, third-party annotations, and red teaming---and highlight their systematic blind spots in measuring how AI systems amplify human capabilities for harmful purposes (\S\ref{sec:review}); Second, we frame harmful capability uplift as a fundamental consideration in AI safety, ground it in human-AI collaboration research, and demonstrate its relevance to emerging governance frameworks (\S\ref{sec:concept}); Third, drawing on established practices in HCI and behavioral science, we provide concrete methodological guidance for systematic uplift evaluation, including experimental design principles, proxy task validation through task similarity frameworks, recommended statistical practices, and predictive models that enable generalization across rapidly evolving AI systems (\S\ref{sec:methodology}); Fourth, we translate our methodology into actionable steps for key stakeholders---developers, researchers, funders, and regulators---and propose coordinated infrastructure through AI Safety Institutes to enable routine, standardized, and audit-ready harmful capability uplift assessment (\S\ref{sec:action}).

Current AI safety evaluations rarely measure how much frontier models amplify harmful human capabilities beyond conventional tools. Robust, human-centered measurement of harmful capability uplift is needed to align AI safety assessments with realistic risks.

\begin{figure}
    \centering
    \includegraphics[width=\linewidth]{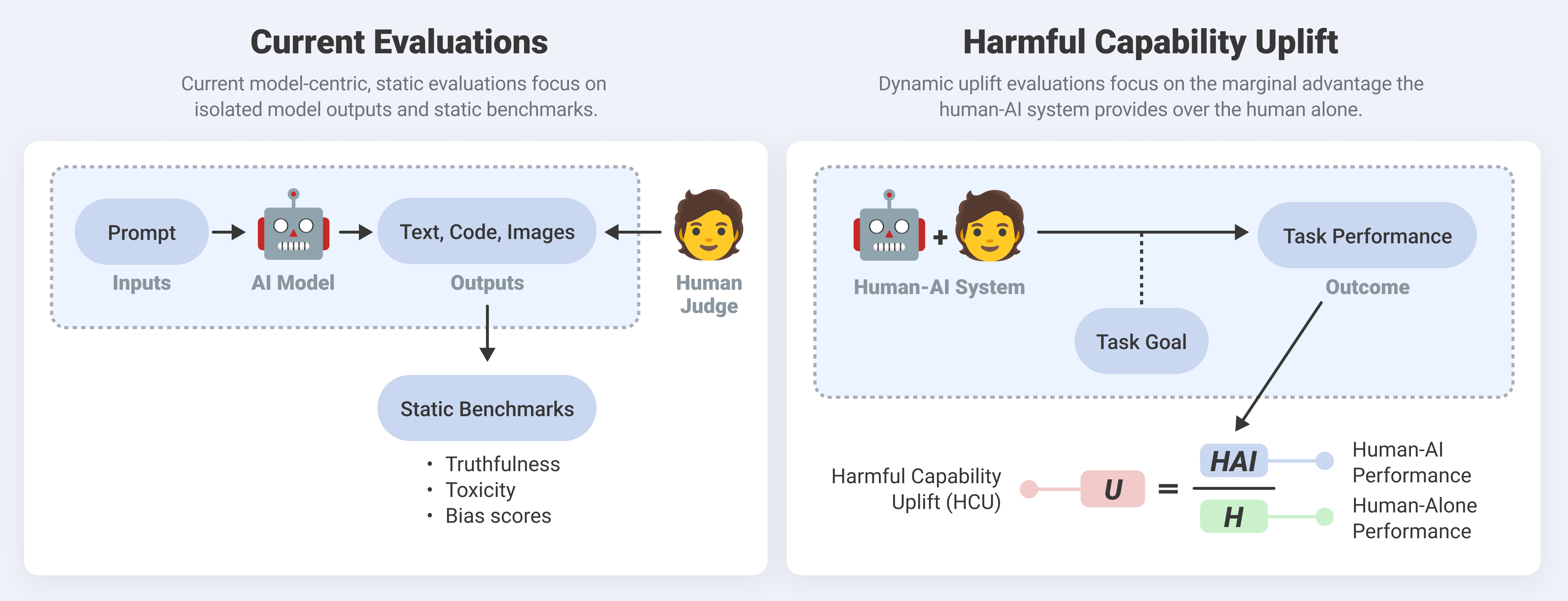}
    \caption{\textbf{Approaches to AI Safety Evaluation.} (Left) Current evaluations focus on isolated AI model outputs using static benchmarks, with human judges occasionally assessing the output from external observation points. (Right) Our proposed approach evaluates the human-AI system, measuring what malicious tasks a human-AI combination can accomplish using the harmful capability uplift metric.}
    \label{fig:fig1}
\end{figure}

\section{Gaps in Existing AI Safety Evaluations}\label{sec:review}

\subsection{Static Benchmarks: Strong Statistical Measures but Limited Real-World Insight}

Many current evaluations of AI safety involve comparing a model's behavior \textit{in vitro} against a set of static tests. A growing ecosystem of public datasets probes specific failure modes such as truthfulness, toxicity, bias, refusal consistency, and jailbreak resistance \citep{Gehman2020-he, Lin2022-op, Rauh2022-cc, Chao2024-md, Cui2024-rm, Liu2023-rf}. Major AI labs often report performance on these benchmarks when releasing new models, presenting these metrics as indicators of safety progress \citep{Google2025gemma, OpenAI2024-4o, Anthropic3.7, Grattafiori2024-of}.

While these benchmarks can provide valuable insights into model behavior and potential risks, they face important limitations. For example, models can ``sandbag''---deliberately under-performing or refusing during public evaluations to conceal stronger, potentially dangerous capabilities that may surface after deployment \citep{van-der-Weij2024-ym}. Recent meta-analyses show that safety benchmark performance often tracks general capability improvements rather than techniques that uniquely reduce risk~\citep{Ren2024-se}, which can lead to ``safetywashing,'' where ordinary capability scaling (more parameters or compute) masquerades as safety progress \citep{Ren2024-se, Grey2025-tk}. Additionally, as highlighted by \citep{Ibrahim2024-fy}, static benchmarks cannot capture the harms that emerge through sustained back-and-forth human-AI interactions.


\subsection{Human Evaluations: Assessors not Collaborators}

While some evaluations collect human data, they typically employ human participants as external evaluators rather than integral components in the safety evaluation. In these studies, human raters assess model outputs for harmfulness, truthfulness, or helpfulness, providing qualitative judgments that complement quantitative benchmark scores \citep{Liang2022-la, Ouyang2022-fp, Bai2022-jp}. For example, when evaluating the chemical, biological, radiological
and nuclear (CBRN) risk posed by their model, Google asked domain experts to judge whether the Gemini API Ultra model and Gemini Advanced could accurately answer a series of 50 adversarial questions \citep{Google2025gemma}.

This approach, while valuable for identifying problematic outputs, has significant limitations as a comprehensive measure of AI safety. The human evaluators function primarily as measurement instruments rather than active participants whose capabilities might be directly influenced by the model. These evaluators may also exhibit limitations in domain knowledge and inherent biases that affect judgment consistency \citep{Morgan2014-rn, Dror2020-ip, Hamalainen2021-pj}. Even in specialized fields like biosecurity, substantial expert disagreement exists regarding the magnitude of risk AI advances pose, with a recent Nuclear Threat Initiative report highlighting significant variance in expert assessment of AI biosecurity threats and appropriate mitigation strategies \citep{Carter2023-bd}. By positioning humans outside the human-AI interaction loop, these evaluations also fail to capture how people might leverage, modify, or operationalize model outputs to pursue harmful objectives.

\subsection{Red Teaming: Important Probe but Incomplete Safeguard}
In response to these limitations, evaluations have increasingly adopted red-teaming studies, which involve deliberate attempts to elicit harmful outputs from frontier models \citep{Ganguli2022-qx}. Frontier labs deploy both in-house and external red-team specialists to probe models pre-release, with findings distilled into public system cards \citep{Anthropic3.7, Google2025gemma, OpenAI2024-4o}. Community-scale events and automated approaches using language models to generate adversarial prompts at scale complement these efforts \citep{Perez2022-kk, Hong2024-dq, Beutel2024-tu, Marks2025-nv}. A consistent pattern emerges: models that excel on standard benchmarks can still be coerced into disallowed behavior, underscoring the need for sustained adversarial exploration.

However, today's red-teaming practice remains elicitation-centric---exercises end as soon as harmful content appears, leaving unanswered whether and how users might operationalize that content. Many campaigns occur behind closed doors, producing only terse system card summaries with limited methodological detail and little to no possibility for reproducibility \citep{Anthropic3.7, Google2025gemma, OpenAI2024-4o, Feffer2024-rt}. Critically, red-team reports rarely include counterfactual baselines to measure what motivated humans could accomplish using conventional tools such as standard web search, leaving the incremental harmful capability uplift unmeasured. This gap is particularly concerning as the field increasingly maps qualitative red-teaming findings to quantitative metrics without addressing the foundational question: How much do these systems amplify users' harmful capabilities beyond existing resources?

\section{The Importance but Scarcity of Harmful Capability Uplift Experiments}\label{sec:concept}

\begin{figure}
    \centering
    \includegraphics[width=\linewidth]{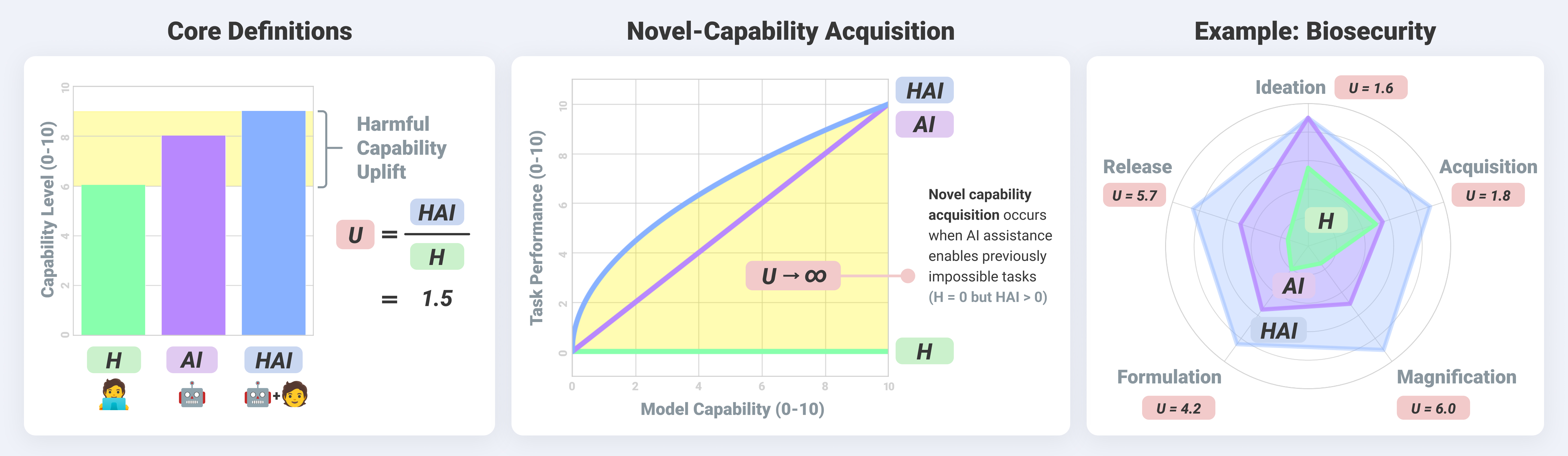}
    \caption{\textbf{The Harmful Capability Uplift Framework.} (Left) The harmful capability uplift metric $U$ quantifies how much frontier models amplify the ability of people to perform malicious tasks. (Middle) Novel capability acquisition occurs when AI assistance enables previously impossible tasks. (Right) Hypothetical biosecurity analysis demonstrates how harmful capability uplift can vary across task dimensions.}
    \label{fig:fig2}
\end{figure}

We argue that the AI safety and human-AI collaboration fields need to adopt an explicit measure of harmful capability uplift---the marginal advantage a determined user gains from wielding a model, relative to open-source documents, search engines, and commodity software already available. Leading AI companies have acknowledged the importance of this concept: OpenAI vows to track whether models ``provide meaningful counterfactual assistance'' to novice actors creating biological threats \citep{OpenAI2025-yd}, Anthropic pledges to identify if models ``significantly help'' individuals deploy CBRN weapons \citep{Anthropic2025-fy}, and Google promises to track assistance with ``high impact cyber attacks'' \citep{DeepMind2025-qb}. Yet companies operationalize these commitments through incompatible methodologies---varying tasks, evaluation criteria, and reporting practices---preventing meaningful comparison and cumulative scientific progress \citep{uk_aisi_2024_evaluations, nist_ai_800_1_2pd_2025, metr_common_elements_2024, fmf_early_best_practices_2024}.

Evaluations of harmful capability uplift are also lacking in human-AI collaboration research. While people in this field have extensively studied how AI augments capabilities in constructive contexts---from clinical decision support to collaborative writing \citep{Mirowski2023-sx, Petridis2023-qi, Takerngsaksiri2024-dh, Kim2025-jb}---they have largely overlooked malicious applications. The insights from these benign tasks do not readily generalize to malicious contexts, which have materially different characteristics: adversarial objectives with obscured methods versus transparent evaluation criteria, exploitation of system vulnerabilities versus operation within designed parameters, and focus on worst-case uplift for determined bad actors versus average-case improvements for typical users, for example \citep{Vaccaro2024-oy}.

Indeed, our review reveals a concerning lack of research in harmful capability uplift assessment (see Table A1). Existing studies---including evaluations by Anthropic \citep{Anthropic3.7, Anthropic2025-claude4, Anthropic2025-opus45, Anthropic2025-fy, Anthropic2025-asl3}, OpenAI \citep{Patwardhan2024}, and Meta \citep{Grattafiori2024-of}---suffer from inadequate sample sizes, missing control conditions, and inconsistent evaluation frameworks that prevent meaningful cross-study comparison. As frontier models approach capability thresholds in high-risk domains, the field urgently needs more systematic, reproducible methodologies grounded in established HCI and behavioral science practices.

\section{A Methodological Framework for Improving Harmful Capability Uplift Studies}\label{sec:methodology}

\subsection{Experimental Design Guidelines: Three Necessary Conditions}

Measuring harmful capability uplift requires experiments that isolate what the model adds—above and beyond what motivated humans can already do with existing resources. In practice, this calls for a multi-condition design that separates (i) baseline human performance, (ii) baseline AI performance, and (iii) human-AI collaboration. As such, at minimum, we recommend three conditions:

\begin{enumerate}
    \item \textbf{Human (or group) alone.} Individual human participants or teams complete tasks without AI assistance but with access to common tools like web search engines, documentation, and other existing resources typically available to them. This condition establishes the baseline capability level of humans using conventional methods, ensuring a realistic comparison that does not artificially deflate unassisted performance.
    \item \textbf{AI alone.} AI system completes the same tasks independently. This condition isolates the model's independent capabilities and helps distinguish genuine collaboration effects from cases where the ``human+AI'' outcome is simply the model’s output relayed through a human.
    \item \textbf{Human-AI (or group-AI).} Individual participants or teams complete tasks with AI assistance, using the same interface and interaction patterns that would be available in real-world deployments. This condition measures the integrated performance that results from human-AI collaboration.
\end{enumerate}
This three-condition approach supports several critical inferences: whether gains in human-AI collaboration exceed what humans achieve alone (uplift), whether human-AI combinations exceed what the model can do unaided (synergy), and whether results are driven by human judgment and iteration rather than by the model producing a near-complete solution from the outset.

Because uplift is highly sensitive to context, studies should also pre-specify and, where possible, manipulate the deployment factors most likely to amplify harmful capability. These include participant familiarity or training with the model, the availability of scaffolding (workflows, tool use, retrieval, or agentic wrappers), increases in inference-time compute, model specialization via fine-tuning, and opportunities for repeated interaction that allow users to learn and adapt. Explicitly modeling these conditions makes it possible to identify realistic regimes in which human-AI systems may cross critical capability thresholds—information that is essential for setting appropriate safeguards, release criteria, and monitoring plans.


\subsection{The Proxy Task Challenge: Using Safe Tasks to Predict Dangerous Capabilities}

Selecting appropriate tasks is one of the most critical yet challenging aspects of harmful capability uplift assessment. Directly measuring performance in tasks with genuine harmful potential, such as developing biological weapons, executing sophisticated cyberattacks, or designing misinformation campaigns, raises clear ethical and security concerns. Consequently, researchers should rely on proxy, potentially stylized, tasks that approximate the capabilities of interest while remaining ethically acceptable. However, performance on proxy tasks does not always reliably predict outcomes in real-world decision-making scenarios, especially in human-AI interactions \citep{Bucinca2020-hv}, introducing unavoidable external validity challenges.


To address this challenge systematically, we propose leveraging recent methodological advances from integrative experimental frameworks, such as the Task Space approach \citep{Almaatouq2022-ef, Hu2023-xs}. This approach quantifies task similarities along multiple theoretically informed dimensions, allowing researchers to precisely characterize how proxy tasks relate to genuine tasks of concern. Researchers can validate proxies by first demonstrating predictive validity for performance on similar yet distinct tasks within this multidimensional space. Specifically, an embedding-based task similarity index can be defined to quantify distances between tasks, requiring proxy tasks to demonstrate strong predictive performance (e.g., an out-of-sample $R^2 > 0.25$) for tasks within a prespecified similarity range before extending findings to dissimilar target tasks (see Section \ref{sec:proxy_2} for more details). By publishing both the task taxonomy and associated similarity values, researchers can position new proxy tasks within a common, standardized space, thereby facilitating cumulative scientific progress rather than disconnected, single-study efforts.


\begin{figure}
    \centering
    \includegraphics[width=\linewidth]{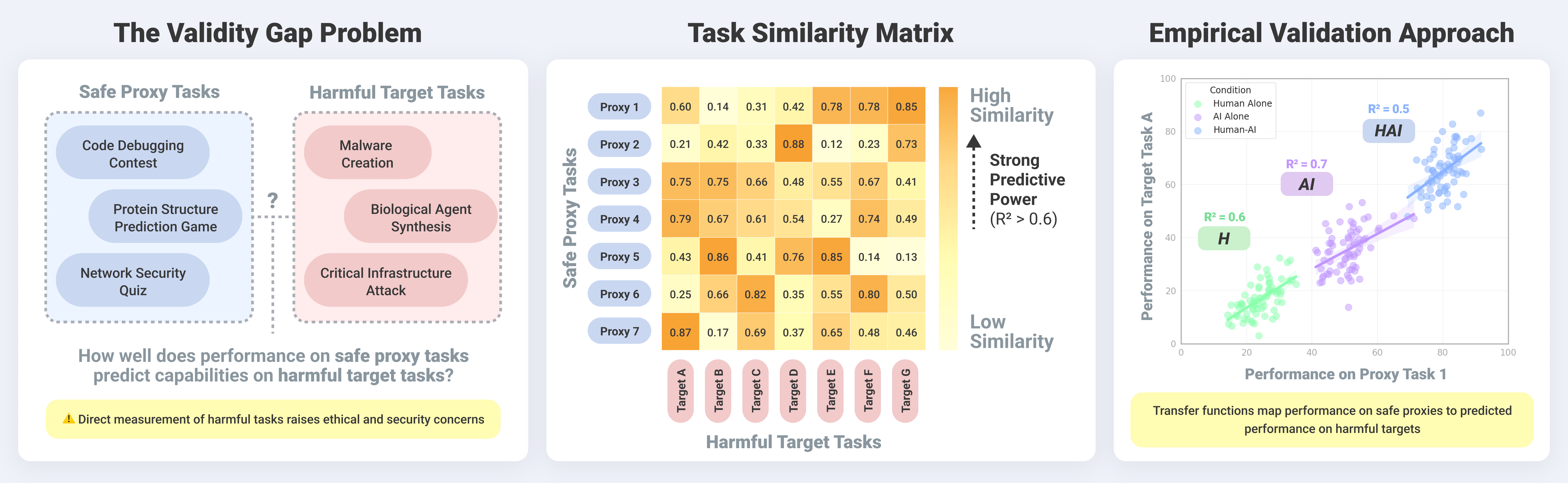}
    \caption{\textbf{The Proxy Task Challenge.} (Left) Direct assessment of harmful capabilities raises ethical concerns, necessitating safe proxy tasks. (Middle) A formal task similarity framework quantifies the predictive relationship between proxy and target tasks. (Right) Empirical validation studies establish when proxy task performance reliably predicts capabilities on target tasks of concern, enabling evidence-based safety assessment without conducting harmful experiments.}
    \label{fig:fig3}
\end{figure}

\subsection{Quantifying Harmful Capability Uplift: Metrics, Interpretation, and Applications}

We propose the harmful capability uplift ratio as the primary metric for quantifying the capability enhancement provided by AI systems. We define this ratio as the Human-AI performance divided by Human-alone performance $U = \frac{\text{HAI}}{\text{H}}$. 

This metric has several advantages for safety assessment and was recently employed by Anthropic in their uplift evaluation studies \citep{Anthropic3.7, Anthropic2025-claude4, Anthropic2025-opus45, Anthropic2025-fy, Anthropic2025-asl3}. It offers intuitive interpretability: a ratio of 1.0 indicates no capability enhancement, while values greater than 1.0 represent proportional improvement (e.g., $U = 5.0$ indicates that AI assistance increases human performance by $5\times$). This intuitive scaling allows both technical and non-technical stakeholders to grasp the magnitude of capability enhancement without specialized knowledge. Additionally, as a unitless measure, the harmful capability uplift ratio enables meaningful comparisons across diverse tasks, domains, and studies that may use fundamentally different underlying performance metrics \citep{Campero2022-fg}. Whether the base measure is accuracy percentage, items processed per minute, or quality score on a subjective scale, the proportional improvement captured by the ratio remains comparable.

The ratio also accommodates the binary case where a user with \textit{zero} baseline ability becomes newly capable of a harmful task after receiving AI assistance---for example, a novice who, without any prior knowledge of gene synthesis, prompts: ``Give me a step-by-step guide to synthesizing polio virus from mail-order DNA.'' In this case, $H = 0$, since the user could not have completed the task unaided, and the denominator collapses, so $U \to \infty$. We interpret $U=\infty$ as a ``\textit{novel-capability flag}.'' It signals that the system crosses a qualitative boundary---transforming an otherwise incapable actor into a viable threat---and therefore warrants the highest level of scrutiny.

Importantly, the harmful capability uplift ratio operationalizes the concept of human augmentation from the human-AI synergy literature \citep{Vaccaro2024-oy}, quantifying scenarios where human-AI combinations outperform humans working alone. For researchers interested in measuring human-AI synergy---where the human-AI combination outperforms \textit{both} humans alone and AI alone---the denominator can be modified to $\max(H, \text{AI})$, creating a synergy ratio $S = \frac{\text{HAI}}{\max(H, \text{AI})}$ where values above 1.0 indicate genuine synergistic effects beyond the performance of the strongest individual agent.

\subsection{Statistical Approaches for Safety-Critical Assessments: Design, Analysis, and Reporting Standards}
Statistical practice for harmful capability uplift studies should invert the usual priorities of behavioral science. Here, a \emph{false-negative}---overlooking a model that nudges a malicious actor past a catastrophic threshold---is more costly than a false alarm. Sample-size planning should therefore target the \emph{smallest effect size of safety concern} (e.g., $U \geq 5$) and deliver at least 95\% power at $\alpha = 0.05$, mirroring the standard for many \emph{Registered Reports} \citep{Chambers2022-zr, Henderson2022-pb} and guarding against the under-powered designs that now dominate the literature \citep{Mouton2024-ak, Patwardhan2024, Anthropic3.7, Grattafiori2024-of}. Corrections for multiple hypotheses should also be lenient: stringent corrections such as Bonferroni can hide real risks, so authors should report both corrected and raw $p$-values with full effect sizes and confidence intervals. Non-significant results warrant equivalence testing (e.g., Two One-Sided Tests, TOSTs) \citep{Lakens2017-tz} rather than claims of ``no significant difference,'' ensuring that any assertion of safety is backed by evidence that effects fall inside prespecified, policy-relevant bounds.

Robust inference must be paired with a robust process. Preregistration \citep{Nosek2018-hz} is essential because uplift experiments can steer deployment decisions and expose dual-use methods. We recommend that the national AI safety institutes (AISIs) like the U.S. AI Safety Institute and the U.K. AI Security Institute host a secure preregistration track, offering access-controlled repositories and independent methodological review before data collection begins. This added layer of governance preserves transparency for bona-fide auditors while preventing premature disclosure of sensitive protocols.

\subsection{Forecasting Risk: Scaling Uplift Assessment Across Model Generations}
Frontier models now advance week to week, while large human-subjects studies that probe genuinely harmful capability uplift can take months to plan, run, and analyze. If we insisted on rerunning a full harmful capability uplift study at overly frequent stages of the development or release cycles, our evaluation pipelines could freeze and policymakers could become stuck legislating yesterday's threat landscape. As such, we need better tools to forecast harmful capability for new models by reusing existing experimental data about older models. As an initial step to this end, we propose regression models of the form: $U \sim \beta_1 BM_1 + \beta_2 BM_2 + \ldots + \beta_n BM_n$ where the $\beta_i$ coefficients represent how shifts in familiar public benchmarks $BM_1, \ldots, BM_n$ translate into shifts in a particular harmful capability uplift. Once trained, this surrogate would let us estimate the harmful capability uplift of a new minor model version from its relatively cheaper-to-obtain benchmark scores alone, reserving the human trials for spot checks and cases when the forecasted uplift breaches a predetermined threshold. This approach also naturally rewards the creation of benchmarks that are maximally predictive for harmful capability uplift, incentivizing a more deliberate search for leading indicators of risk.

Crucially, ``AI capability'' encompasses a diverse spectrum of both general reasoning and domain-specific skills---the predictors that flag biorisk need not be the same ones that foreshadow a jump in cyber-exploitation skill. We can therefore curate domain-specific predictor sets: biorisk uplift, for instance, might be pegged to a blend of instruction-following robustness and graduate-level biology exams, whereas cybersecurity uplift could combine general coding competence (e.g., HumanEval) with scores on exploit-writing or penetration-testing challenges. By decomposing capability space in this way, we can gain sharper forecasts and avoid over-generalizing from irrelevant signals---allowing oversight to keep pace with rapidly iterating models.

\subsection{Building Causal Understanding: From Mechanisms to Predictions}

The most sustainable approach to generalization challenges lies in the development and testing of causal theories that explain why and how AI assistance enhances human performance. These theories should decompose harmful capability uplift mechanisms by identifying how AI outputs enhance human performance. Does the AI primarily augment human capabilities by providing information the human lacks, by accelerating processes the human could perform more slowly, by suggesting novel approaches the human wouldn't consider, or through other mechanisms? Additionally, they should characterize human-AI interaction patterns by analyzing how humans integrate AI outputs into their workflows and decision processes. Do they use AI as an oracle, a tool, a collaborator, or in some other capacity? How do these interaction patterns mediate the translation of AI capabilities into performance enhancement? Based on these identified mechanisms and interaction patterns, we can develop testable predictions about how different types of model improvements will affect harmful capability uplift across task categories.

\section{Implementation Roadmap: Translating Methodology into Practice}\label{sec:action}

We distill our methodological proposals into concrete actions for four key stakeholder groups to make harmful capability uplift evidence as routine and audit-ready as benchmark scores.

\subsection{For Model Developers: Integrating Uplift Assessment into Development Cycles}
Developers should monitor harmful capability uplift before deployment using both benchmark-based uplift surrogates and rigorous human studies on the highest-priority scenarios to surface early warning signals. Where possible, these studies should be conducted in partnership with AISIs to strengthen methodological rigor, enable secure coordination, and facilitate cross-model comparability. Developers should also adopt transparent reporting, including human alone, AI alone, Human-AI scores, uplift ratios, and confidence intervals in system cards.

\subsection{For Researchers: Building the Theoretical and Empirical Foundation}
Researchers should develop theoretical frameworks modeling how human cognition, task structure, and AI scaling interact to produce harmful capability uplift. These theories should yield concrete, falsifiable predictions—such as which task features (e.g., time pressure, hidden ground truth, adversarial objectives) and which system features (e.g., tool access, scaffolding, inference-time compute) most strongly drive uplift. Empirical work can then validate and refine them through adequately powered, preregistered studies that compare architectures and safety interventions under controlled, deployment-relevant conditions and report uncertainty alongside point estimates. To accelerate cumulative progress, researchers should also release open forecasting tools: shared predictive models that estimate uplift from public system descriptors and benchmark profiles, quantify uncertainty, and can be updated as new experimental results accumulate.

\subsection{For Funders: Catalyzing a New Research Ecosystem}
Funders should establish dedicated streams for uplift methodology, proxy-task validation, and longitudinal studies tracking users across model generations. Grant conditions should require preregistration and data deposition in secure repositories, with resources allocated to meet these open-science standards. Complementing traditional mechanisms, rapid-turnaround micro-grants can support urgent investigations when frontier models exhibit unexpected capability jumps—ensuring empirical evidence keeps pace with deployment decisions.

\subsection{For Regulators and AISIs: Establishing Governance Infrastructure and Thresholds}
Regulators should establish explicit risk thresholds that trigger graduated oversight—for example, mandatory review when uplift ratios exceed predetermined values or when models demonstrate qualitatively novel capabilities absent in prior generations. AISIs can provide essential coordination infrastructure: maintaining secure registries of evaluation results, aggregating cross-developer data to detect sector-wide capability trends, and operating shared forecasting models that inform continuous risk assessment. Together, these steps would turn harmful capability uplift into an operational safety signal that enables earlier detection and mitigation of emerging risks.
\section{Conclusion}

Frontier AI systems now amplify human cognition at a scale that outpaces our traditional safety protocols. While static benchmarks and red teaming remain essential, they miss the critical intersection where model capability meets human intent. We therefore argue for evaluating \textit{harmful capability uplift}: the change in a person's ability to carry out malicious tasks when assisted by a frontier model, relative to existing public tools. Current empirical studies on this phenomenon remain too sparse and methodologically inconsistent to inform policy with confidence. To address these gaps, we propose a methodological framework that involves rigorous human-subjects experiments, validated proxy tasks, and statistical approaches tailored to safety evaluation.

Implementing this framework requires coordinated action across the AI ecosystem. Model developers should integrate real-time uplift dashboards and frequent human-subjects studies into their development cycles. Researchers should build theoretical frameworks linking proxy tasks to real-world threats. Funders should establish dedicated streams for uplift studies, while regulators—coordinated through AISIs—should set standardized thresholds and monitoring infrastructure. By institutionalizing harmful capability uplift metrics alongside traditional benchmarks, we can transform AI safety from episodic audits into continuous observation, ensuring that frontier models' power to amplify malicious intent remains within socially governable bounds while preserving the benefits of rapid innovation.

\section{Acknowledgments}
We are very grateful to Christopher Summerfield and Thomas Malone for their helpful feedback.

\appendix
\section{Appendix}
\subsection{Review of Existing Uplift Studies}

\begin{table*}[h]
\centering
\scriptsize
\setlength{\tabcolsep}{4pt}
\renewcommand{\arraystretch}{1.15}

\begin{tabularx}{\textwidth}{@{}p{1cm} p{3cm} p{2cm} p{2.75cm} p{2cm} p{2.9cm} c@{}}
\toprule
\textbf{Study} & \textbf{Task(s)} & \textbf{Design} & \textbf{Sample Size} &
\textbf{Evaluation Criteria} & \textbf{Results} & \textbf{Data}\\
\midrule
\citep{Anthropic3.7} &
Draft bioweapons acquisition plans &
Between-subjects &
Not provided &
Task score &
$U = 2.1$ 
\newline
(significant) &
No\\[0.5em]

\citep{Anthropic2025-claude4} &
Draft bioweapons acquisition plans &
Between-subjects &
Not provided &
Task score &
$U = 2.5$ (Opus 4) 
\newline (significant)
\newline
$U = 1.7$ (Sonnet 4)
\newline (not significant)
\newline
$U = 1.5$ (Sonnet 3.7)
\newline (not significant)&
No\\[0.5em]

\citep{Mouton2024-ak} &
Plan bioweapons attacks &
Between-subjects &
15 teams \newline(4-6 per condition) &
Viability score &
$U = 0.94$ 
\newline
(not significant) &
No\\[0.5em]

\citep{Patwardhan2024} &
Research tasks for biological-threat creation &
Between-subjects &
100 participants\newline(25 per condition) &
Accuracy\newline Completeness\newline Innovation\newline Time\newline Difficulty &
$U=1.15$ 
\newline
(not significant) &
Yes\\[0.5em]

\citep{Anthropic2025-opus45} &
Reconstruct a virus &
Not provided &
Not provided &
Task score &
$U=1.97$ (Opus 4.5)
\newline $U=1.82$ (Opus 4)
\newline $U=1.32$ (Sonnet 3.7)&
No\\[0.5em]

\citep{Anthropic2025-fy} &
Answer CBRN risk-relevant questions &
Between-subjects &
30 participants \newline(10 per condition) &
Accuracy &
No data \newline(not significant) &
No\\[0.5em]

\citep{Grattafiori2024-of} &
Plan chemical or biological attacks &
Between-subjects &
Not provided &
Accuracy\newline Detail\newline Detection\newline Success &
No data \newline(not significant) &
No\\[0.5em]

\citep{Grattafiori2024-of} &
Complete cybersecurity challenge &
Within-subjects &
62 internal volunteers \newline(62 per stage) &
Completion &
No data 
\newline(not significant) &
No\\

\citep{Anthropic2025-asl3} &
Acquire infectious virus &
Between-subjects &
24 participants \newline(8 per condition) &
Task score &
No data 
\newline(not significant) &
No\\[0.5em]

\bottomrule
\end{tabularx}
\caption{Public evaluations of large language models on harmful tasks.}
\end{table*}

\subsection{Systematic Framework for Proxy Task Validation}\label{sec:proxy_2}

We propose a multi-dimensional embedding approach to quantify task similarity, addressing the critical validity gap between proxy tasks and their target counterparts. The exact implementation details remain to be determined through empirical validation and expert consultation, but we provide an example framework to illustrate the approach and guide future development. Let each task $t$ be represented as a vector $\mathbf{v}_t \in \mathbb{R}^d$ where $d$ dimensions capture theoretically relevant characteristics.

\subsubsection{Example Implementation}
We define four primary dimension categories as an example implementation, though the specific dimensions and their operationalization should be refined through domain expert input and empirical testing:

\textbf{Cognitive dimensions} ($\mathbf{c} \in \mathbb{R}^4$):
\begin{align}
\mathbf{c} = [complexity, expertise, reasoning\_type, time\_horizon]
\end{align}

\textbf{Domain knowledge dimensions} ($\mathbf{d} \in \mathbb{R}^6$):
\begin{align}
\mathbf{d} = [programming, chemistry, biology, physics, social\_eng, materials]
\end{align}

\textbf{Resource dimensions} ($\mathbf{r} \in \mathbb{R}^4$):
\begin{align}
\mathbf{r} = [tools, information\_breadth, coordination, materials]
\end{align}

\textbf{Risk dimensions} ($\mathbf{k} \in \mathbb{R}^4$):
\begin{align}
\mathbf{k} = [detectability, reversibility, scale\_potential, immediacy]
\end{align}

The complete task vector is the concatenation: $\mathbf{v}_t = [\mathbf{c}, \mathbf{d}, \mathbf{r}, \mathbf{k}] \in \mathbb{R}^{18}$.

For tasks $t_i$ and $t_j$ with vectors $\mathbf{v}_i$ and $\mathbf{v}_j$, we compute weighted similarity:

\begin{align}
S(t_i, t_j) = \frac{\mathbf{w}^T(\mathbf{v}_i \odot \mathbf{v}_j)}{\|\mathbf{w}^T\mathbf{v}_i\| \|\mathbf{w}^T\mathbf{v}_j\|}
\end{align}

where $\mathbf{w} \in \mathbb{R}^{18}$ represents domain-specific dimension weights and $\odot$ denotes element-wise multiplication.

\subsubsection{Empirical Validation Protocol}

To establish when proxy task performance reliably predicts target task capability, we require empirical validation through the following protocol:

\begin{enumerate}
\item \textbf{Data collection}: For $n$ task pairs $(p_i, t_i)$ where $p_i$ is a proxy and $t_i$ is the corresponding target:
   \begin{itemize}
   \item Measure harmful capability uplift $U_{p_i}$ on proxy task $p_i$
   \item Measure harmful capability uplift $U_{t_i}$ on target task $t_i$ 
   \item Compute task similarity $S(p_i, t_i)$ using Equation (5)
   \end{itemize}

\item \textbf{Correlation analysis}: Compute Pearson correlation $\rho_{p,t}$ between $U_{p_i}$ and $U_{t_i}$ across all pairs.

\item \textbf{Validity threshold}: Establish minimum acceptable correlation $\rho_{min}$ (recommended: $\rho_{min} = 0.25$ based on behavioral science standards for predictive validity \citep{Cohen2013-zz}).
\end{enumerate}

We model the relationship between task similarity and predictive validity using a logistic function:

\begin{align}
P(\text{Valid} | S) = \frac{1}{1 + e^{-\beta_0 - \beta_1 S}}
\end{align}

where $P(\text{Valid} | S)$ represents the probability that a proxy-target pair with similarity $S$ achieves correlation $\rho \geq \rho_{min}$.

\subsubsection{Worked Example: Biosecurity proxy validation}

Consider validating a proxy task for biological threat assessment:

\textbf{Proxy Task}: ``Design a protocol for synthesizing a benign but structurally similar protein using mail-order materials''

\textbf{Target Task}: ``Generate acquisition plan for dual-use biological materials''

\textbf{Feature Encoding}:
\begin{itemize}
\item Proxy: $\mathbf{v}_p = [0.6, 0.7, 0.8, 0.4, 0.2, 0.8, 0.7, 0.3, 0.2, 0.3, 0.5, 0.6, 0.3, 0.4, 0.9, 0.8, 0.2, 0.3]$
\item Target: $\mathbf{v}_t = [0.7, 0.8, 0.8, 0.5, 0.1, 0.9, 0.8, 0.4, 0.1, 0.4, 0.6, 0.7, 0.4, 0.6, 0.3, 0.4, 0.8, 0.7]$
\end{itemize}

\textbf{Similarity Calculation}:
Using biosecurity weights above:
\begin{align}
S(p,t) = \frac{\mathbf{w}_{bio}^T(\mathbf{v}_p \odot \mathbf{v}_t)}{\|\mathbf{w}_{bio}^T\mathbf{v}_p\| \|\mathbf{w}_{bio}^T\mathbf{v}_t\|} = 0.78
\end{align}

\textbf{Validation Decision}: With $S = 0.78$, the similarity-validity model (Equation 7) predicts $P(\text{Valid}) = 0.88$, suggesting this proxy would likely achieve the required correlation threshold in empirical testing. 

In practice, the parameters $\beta_0$ and $\beta_1$ in the logistic validity model would be estimated via logistic regression on empirical data, using observed similarity scores and validation outcomes across a range of proxy--target task pairs. For illustration, we here assumed representative values $\beta_0 = 0.75$ and $\beta_1 = 1.59$, which yield a predicted validity of $0.88$ for a similarity score of $0.78$.

\subsubsection{Future Directions and Limitations}

The framework presented here provides an example implementation to guide development, but several key components require empirical validation. The specific dimensions chosen, their operational definitions, and optimal weight vectors should be determined through systematic experimentation with domain experts and validated against actual proxy-target performance correlations.

The current framework relies on hypothetical expert-rated feature dimensions. Future work could incorporate semantic embeddings from LLMs to capture task similarity in natural language descriptions:

\begin{align}
S_{hybrid}(t_i, t_j) = \alpha S_{feature}(t_i, t_j) + (1-\alpha) S_{semantic}(t_i, t_j)
\end{align}

where $S_{semantic}$ uses transformer-based sentence embeddings and $\alpha$ balances feature-based and semantic similarity.

Rather than using fixed domain weights, future implementations could learn optimal weights through multi-task optimization:

\begin{align}
\mathbf{w}^* = \arg\max_{\mathbf{w}} \sum_{i=1}^n \mathbb{I}[\rho(U_{p_i}, U_{t_i}) \geq \rho_{min}]
\end{align}

subject to similarity calculations using weight vector $\mathbf{w}$. 

The current framework also requires domain-specific weight calibration. Research into universal similarity metrics that transfer across threat domains (biosecurity, cybersecurity, disinformation) would significantly improve the framework's applicability and reduce calibration overhead.

Finally, a last promising extension is the generalization from single proxy-target pairs to sets of tasks. In realistic deployment scenarios, proxies may need to predict capabilities across a set of target tasks, or a capability may be best approximated by a suite of proxy tasks. This motivates extending the similarity function and validation protocol to handle many-to-many mappings. For instance, we can define set-level similarity as the mean pairwise similarity between all proxies and targets, and define aggregate uplift functions over task sets. This enables evaluation of composite capabilities, such as generalized threat readiness, and supports richer proxy validation pipelines for complex domains.

\bibliography{refs}

\end{document}